\documentstyle{5493_l-aa}
\thesaurus{3(11.19.2;11.09.4;09.04.1;09.08.1)}
\newcommand \be{\begin{equation}}
\newcommand \ee{\end{equation}}
\newcommand \hr{H\,{\sc ii} region}
\newcommand \hrs{H\,{\sc ii} regions}
\newcommand \tv{\tau_{\rm V}}
\begin{document}
\title{Dust extinction of \hrs~ in NGC 598 and NGC 5457
\thanks{Based on observations collected at the Nordic Optical Telescope,
 La Palma, Spain}}
\author{L. Petersen \and P. Gammelgaard}
\offprints{larsp@obs.aau.dk}
\institute{Institut for Fysik og Astronomi, Aarhus Universitet, Ny
 Munkegade, DK-8000 \AA rhus C, Denmark}
\date{Received \hskip 2cm; Accepted}
\maketitle
\begin{abstract}
The dust extinction towards bright \hrs in NGC 598 and NGC 5457 has been 
studied in detail by forming line ratios of Balmer and Paschen emission lines
covering a large wavelength range. Three homogeneous models of the geometrical
distribution
of the emitting sources and the obscuring dust have been tested. Only for low
extinctions can the data be fit by a homogeneous slab. For most of the
observed \hrs~ the Witt et al. (\cite{WTC}) 'dusty nucleus' model matches 
the observations equally well as the usual assumption of a foreground screen,
but the former implies much larger actual dust contents. Spatial variations
in the optical depth in V of the order 0.3 across a region are found.

\end{abstract}
\section{Introduction}
Investigations of external galaxies are often faced with the problem of
disentangling the effects of dust and the physical properties, e.g. age, IMF,
chemical composition etc. of the emitting source.

The effect of dust extinction goes roughly as $\lambda^{-1}$ and will
consequently change the observed spectral energy distribution. Studies
and interpretation of the physical properties in external galaxies
therefore rely on how well we can account for the attenuation of the emitted
light due to dust.
The wavelength dependence makes multiwavelength analyses
favourable and a growing number of investigations combining optical and IR
wavelengths have been made both using imaging (e.g. Jansen et al. \cite{jan};
Witt et al. \cite{wi94}; Evans \cite{ev}; van Driel et al. \cite{driel}) and
spectroscopy (e.g. Puxley \& Brand \cite{pu:br}; Calzetti et al.
\cite{CKS96}).

In the analysis of spectra of extragalactic \hrs~ the correction for dust
extinction is usually made on the assumption that the intrinsic hydrogen
emission line spectrum is known through case B recombination and any difference
between the emergent and intrinsic flux ratios can be ascribed to dust
obscuration following the Galactic extinction curve. 

Throughout the years, the use of the ratio of the two strongest Balmer lines,
H$\alpha$/H$\beta$, sometimes in combination with H$\gamma$, has been
prevailing, albeit they only span a limited wavelength range,
because they will be included in almost any optical spectra and are easy to
detect.
It has long been proposed to expand the baseline for extinction
determination by comparing near-IR hydrogen Paschen lines with emission lines
from the Balmer series at short wavelengths (cf. Greve et al. \cite{gr89};
Osterbrock \cite{os89}).
Such line pairs are all separated by wide wavelength intervals over
which the extinction has a large effect and is therefore potentially easier to
deduce. Additionally, some line pairs form corresponding multiplet lines,
${\rm P}n / {\rm H}n$
originating at the same upper atomic level with relative strength depending
primarily on the transition probability and minimised dependence on theoretical
recombination line calculations (Greve et al. \cite{gr94}).

In Petersen \& Gammelgaard (\cite{pe:ga}; hereafter Paper I)
we have demonstrated the feasibility of observing lines from the Paschen 
series and the short end of the Balmer series from giant extragalactic \hrs~
simultaneously with a two-channel spectrograph. In the present paper we will
further discuss dust extinction towards \hrs~ in NGC 598 and NGC 5457 as derived
from spectra obtained with the same instrument.

The straightforward application of the Galactic interstellar extinction curve
derived from point sources to extended emission regions in external galaxies is,
however, problematic, because this procedure implicitly assumes a dust
configuration of a homogeneous foreground absorbing screen. Intuitively one
would
expect things to be more complicated in reality with dust distributed in a
patchy or clumpy way maybe mixed with the emitting source (Witt et al.
\cite{WTC}; hereafter WTC; Calzetti et al. \cite{cal}, \cite{CKS96};
Puxley \& Brand \cite{pu:br}) 
and with scattering of, particularly blue, photons into the line of sight by
dust grains (Bruzual et al. \cite{bru}; WTC). This could
all lead to a much greyer effective extinction curve than the standard
Galactic curve.
This was exemplified in Paper I by using an extinction curve with $R_{\rm V} =
A_{\rm V} / E_{\rm B - V} = 6$, here we will address this issue by fitting
three different models of relative spatial distribution of emitting source and
dust to the measured line ratios.

\section{The models}
When data from a wide wavelength range are available it is possible to test
various models for the geometrical distribution of the dust. Although the data
presented here only reach P$\delta$ at 1.0~$\mu$m and in two cases P$\gamma$ at
1.1~$\mu$m, they can still be used to test some basic models and have the
advantage of being recorded simultaneously with the same instrument as the
Balmer emission lines. Three models have been considered: (1) The naive but
widely used uniform foreground extinction screen, (2) a slab model of
homogeneous mixture of dust and emitting sources and (3) that of the WTC
scenarios which best models an extragalactic \hr, the so-called "dusty nucleus".

The extinction can be quantified in many ways, e.g. as the logarithmic
extinction at H$\beta$, $C({\rm H}\beta)$, the visual extinction in magnitudes,
$A_{\rm V}$ or as the optical depth in V, $\tv$, depending on the purpose it
is used for. In this work we will express the derived dust
extinction as $\tv$ for uniformity among the models.
In the case of a foreground screen simple relations exist between these, with
$\tv = 1.98 C({\rm H}\beta) =  A_{\rm V} / 1.086$. 

Depending on the actual distribution the same total amount of dust can produce
very
different attenuation of the emitted light. Therefore we define an effective
optical depth $\tau_{\rm eff} = -\ln({\cal F}_{\rm obs} / {\cal F}_{\rm em})$.
In most cases the actual optical depth will exceed $\tau_{\rm eff}$, thus
some source distributions are capable of hiding away larges amount of
extinguishing matter.

More complicated models with clumpiness could be included, but due to the lack
of near-IR lines at longer wavelengths and the rather low extinction for some
of the regions we do not find it justified to attempt to discriminate between
clumpy and homogeneous distributions. Thus we limit ourselves to the three
homogeneous models but it must be stressed that these will produce lower limits
to the dust content as compared to corresponding clumpy models. Calculations of
the effect on the emitted radiation field by a clumpy screen with
dust-free interclump medium excluding scattering (Natta \& Panagia 
\cite{na:pa}), a two-phase slab (Boiss\'e \cite{bos}) and spherical two-phase 
clumpy systems with non-conservative scattering (Witt \& Gordon \cite{wi96})
all show reduced overall
effective optical depth relative to homogeneous distributions of the same mass
and flatter effective extinctions curves.
As clumps will become optically thick much quicker than an equivalent
homogeneous distribution would, a clumpy dust distribution will produce less
reddening in a given geometry with an extended source distribution and approach
lower finite limiting reddening with increasing dust mass (Witt \& Gordon
\cite{wi96}).
Thus the dust content derived on the assumption of homogeneous distributions
can underestimate the true dust mass.

\subsection{Foreground screen}
This is the most simple geometry of dust distribution, inherited from stellar
astronomy but likely to be invalid for many extragalactic studies where more 
complicated geometries are expected (WTC). On the other hand
Calzetti et al. (\cite{CKS96}) found evidence that it might be applicable for
star burst environments, where large fractions of the dust have been depleted
or ejected.
By chance it happens to be the geometry
where any amount of dust has the strongest effect on the emitted flux and taken
at face value it will always yield a lower limit to the true dust content.

For any pair of emission lines the relationship of observed line flux ratio
and the predicted ratio in the absence of dust is
\be
\frac{R_{\rm o}}{R_{\rm p}} =
{\rm e}^{\tau_{\lambda 2}-\tau_{\lambda 1}}.
\ee
Hence, as an exception, the foreground screen has by definition $\tau =
\tau_{\rm eff}$. 
Introducing a normalised extinction curve $\tau_{\lambda} / \tv = 
A_{\lambda} / A_{\rm V}$ one can solve for $\tv$
\be
\tv = \frac {\ln(R_{\rm o} / R_{\rm p})}{\tau_{\lambda 2} / \tv -
\tau_{\lambda 1} / \tv}~ {\rm or} ~
A_{\rm V} = \frac {2.5\log(R_{\rm o} / R_{\rm p})}{A_{\lambda 2} / A_{\rm V} -
A_{\lambda 1} / A_{\rm V}},
\ee
if transforming to magnitude.  
The first expression in Eq. (2) is of cause the motivation for illustrating
the data as in Fig. \ref{N5461}--\ref{N598}. 

\subsection{Homogeneous slab}
A slab model of mixture of dust and emitting sources may resemble the actual
geometry better than the overlying screen in some cases, although a single or
a few knots of emission surrounded by non-uniform dust is probably the
dominating picture. For a homogeneous mixture throughout neglecting scattering
$R_{\rm o} / R_{\rm p}$ can be described in terms of the optical depth
(Disney et al. \cite{dis}, Puxley \& Brand \cite{pu:br})
\be
\frac{R_{\rm o}}{R_{\rm p}} = \frac{\tau_{\lambda 2}}{\tau_{\lambda 1}} \left(
\frac{ 1 - {\rm e}^{-\tau_{\lambda 1}}} {1-{\rm e}^{-\tau_{\lambda 2}}}
\right)
\ee
When introducing relative values $\tau_{\lambda} / \tv$ we end up with
the only free parameter $\tv$, which is fit to give the best match to the
data points.

\subsection{Dusty nucleus}

This is probably the most realistic model of the real dust distribution
affecting the emitted flux from an \hr. The WTC 'dusty galactic nucleus'
environment consists of a dust-free spherical source surrounded by a dust-shell
without emitting sources and is characterised by the optical thickness
of the shell, $\tv$. The main difference between this model and the other two is
the presence of a dust-shell subtending a large solid angle as seen by the
source and the inclusion of scattered light back into the line of sight from it,
whereas the other models only
include pure extinction (absorption and scattering out of the line of sight).
This difference has increasing impact in the short-wavelength end resulting in
a 'bluing' of the spectrum partly cancelling the reddening effect of the 
extinction and leading to a flatter effective extinction curve. Consequently,
the observational effect will be to see values of $\ln(R_{\rm o} / R_{\rm p})$
close to
zero in the blue end of Figs. \ref{N5461}--\ref{N598} even at large optical
depths in contrast to the two other distributions, which lead to decreasing
values of $\ln(R_{\rm o} / R_{\rm p})$ as function of $\tau_{\lambda} / \tv$
with a slope depending on $\tv$.

WTC tabulate the amount of direct and scattered
light received at the central wavelength of the broad band colours calculated
from Monto Carlo simulations of the transfer of radiation. In this analysis
we have interpolated their tables to the wavelength of the relevant emission
lines by cubic splines and use the sum of the two contributions as the observed
flux.

\section{Observations}
The spectra were obtained with the \AA rhus-Troms\o{} Low Dispersion 
Spectrograph at the Nordic Optical Telescope, La Palma on July 31 and August 1
1994, together with some of the data presented in Paper I. 

\begin{table}
\caption[]{\label{obs}Journal of observations}
\begin{flushleft}
\begin{tabular}{llllc}
\noalign{\smallskip}
\hline
\noalign{\smallskip}
\hfil Date & \hfil Galaxy & \hfil H{\sc\,ii} reg.$^{\rm a}$ & Apertures &
Total exp. \\
\noalign{\smallskip}
& & & \multicolumn{1}{c}{(\arcsec)} & (s) \\
\noalign{\smallskip}
\hline
\noalign{\smallskip}
Aug 1  & NGC 598 & N595& $2.2\times33.4$ & 3600 \\
Aug 1  & NGC 598 & B88 & $2.2\times8.3$ & 3600 \\
 & & & $2.2\times8.3$ & 3600 \\
Aug 1  & NGC 598 & N604 & $2.2\times9.2$ & 1800 \\
 & & & $2.2\times7.5$ & 1800 \\
 & & & $2.2\times8.3$ & 1800 \\
Jul 31 & NGC 5457 & N5461 & $2.2\times15.8$ & 3600 \\
Jul 31 & NGC 5457 & H949 & $2.2\times10.8$ & 1800 \\
                     & & H961 & $2.2\times8.3$ & 1800 \\
Aug 1  & NGC 5457 & H1026 & $2.2\times8.3 $ & 6000 \\
 & & N5461 & $2.2\times10.0$ & 6000 \\
 & & H1117 & $2.2\times9.2$ & 6000 \\
\noalign{\smallskip}
\hline
\noalign{\smallskip}
\end{tabular}
\begin{list}{}{}
\item[$^{\rm a}$] ID numbers from Boulesteix et al. (\cite{B74}) and Hodge et
al. (\cite{H90}) are preceded by B and H, respectively
\end{list}

\end{flushleft}
\end{table}

Three extended \hrs~ in NGC 598 all classified as extremely strong by
Boulesteix et al. (\cite {B74}) were observed. 
NGC 604 is the largest \hr~in NGC 598
and resembles in some respect the 30 Dor complex in the LMC. The elemental 
abundance has been studied in detail by D\'\i az et al. (\cite{di87}) using
multiple slit positions. We have positioned the slit in an E-W direction
through the center of the complex and extracted spectra from three bright knots
roughly coinciding with their areas denoted D, B and C. 
In NGC 595, another large region, the slit was positioned across the
southern part. B88 shows a feeble ring-like structure and one
spectrum on the east and west side of the ring was extracted.

The five \hrs~ observed in NGC 5457 are all situated in the south-east part
of the galactic disk.   
The giant complex NGC 5461 was observed in two slit positions both
centered visually on the peak luminosity from short H$\alpha$-exposures. One
position (Jul 31) is E-W while the other is roughly in the NE-SW direction
to include H1026 and H1117 with NGC5461. 
A complete log of the observations is given in Table \ref{obs}. 

During both nights the spectrophotometric
standard star BD $+17\degr 4708$ (Oke \& Gunn \cite{ok:gu}) was observed
multiple times. For all \hrs~ and the standard star the
blue and near-IR spectral ranges were observed simultaneously in the blue and
red channel of the spectrograph to ensure precise relative spectrophotometry
of the Balmer and Paschen lines. The two channels have dispersions of 2.0 
\AA /pixel and 7.5 \AA /pixel and cover 3950--4900 \AA~ and 8000--11000 \AA,
respectively. A detailed description of the instrument and
observation and reduction procedures can be found in Paper I.

\subsection{Data reduction}
All the CCD exposures were reduced with standard IRAF packages for flat
fielding, calibrations and extraction of the one-dimensional spectra.
Rectification of the two-dimensional frames and transformation to a common
spatial coordinate system were done with a special slit mask as explained in
Paper I.

In the combined object frames extraction apertures were set to enclose each
distinct emission region; the resulting apertures are listed in Table \ref{obs}.
The sky background was sampled along large part of the slit, avoiding any
sign of emission, to get a good S/N of the sky subtraction in the near-IR.

\begin{figure*}
\vspace{10.5 cm}
\includegraphics{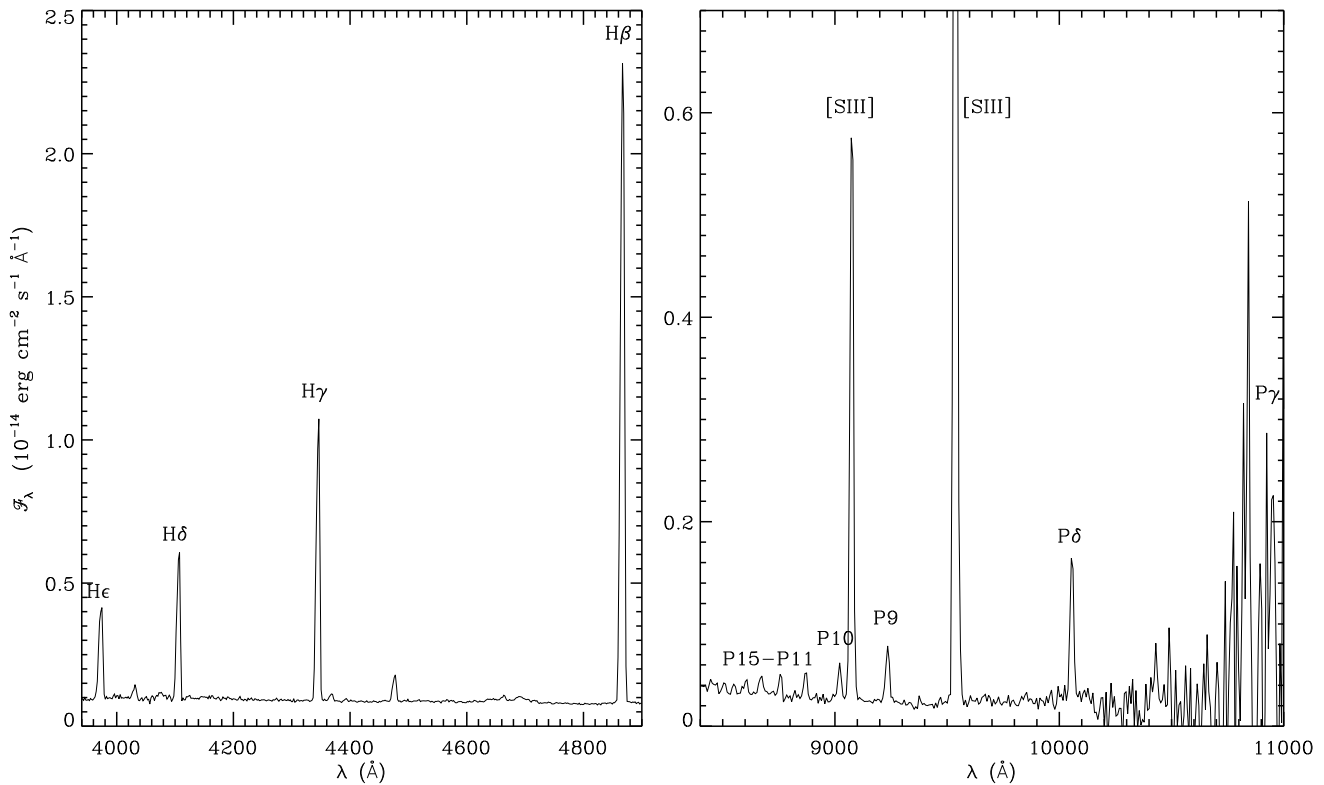}
\caption[]{\label{spec} Left panel, Blue spectrum of NGC 5461 showing four
strong Balmer emission lines. Right panel, near-IR spectrum of the same region.
The Paschen lines P15--P$\gamma$ are clearly identified, although with rapidly
increasing noise for $\lambda > 10500$ \AA. Note the different scales on the
axes}
\end{figure*}

\subsection{Relative line fluxes}
In Fig. \ref{spec} we present an example of the final flux calibrated
spectra in the
blue and near-IR wavelength range from of one of the \hrs~ with high signal. 
The line fluxes have been measured by fitting Gaussian profiles to the line
profiles or two Gaussians when deblending H$\epsilon$ from the nearby
[Ne{\sc\,iii}] line. As evident from Fig. \ref{spec} the spectrum has low S/N
at the position of P$\gamma$. This line was only detected in two of the spectra
and the flux could not
in any case be determined from the IRAF procedure. Instead the flux was measured
by fitting a Gaussian with fixed $\lambda_0$.

For the strong Balmer lines we estimate the accuracy of the flux
measurements to be of the order 5 \%. Because H$\epsilon$ is measured from a
deblending procedure it has a larger flux of error estimated to 20 \%.
In the near-IR flux uncertainties
are dominated by the more critical sky subtraction and determination of the
spectral continuum level. At P15 we tend to reach the series limit where the
faint Paschen lines start to blend giving rise to a false continuum, this line
is therefore excluded from the present analysis. Judged upon the line profile
fitting procedure we assign an error estimate of 20 $\%$ to the two faintest
lines
included (P14 and P13) and to P9, which is strongly affected by the atmospheric
H$_2$O absorption band, and half of this to P12--P10 and P$\delta$.

The nebular Balmer emission lines are affected by the absorption lines in the
underlying stellar spectrum. This will suppress the observed flux of the lines
and relative more so at the short end of the line series, thus partly
counteracting the effect of scattering on dust grains. The exact correction for
underlying Balmer absorption is very uncertain, because it depends on the
actual stellar population and age. Due to the lack of any detailed knowledge
we have decided to use the average absorption equivalent width of 1.9 \AA ,
found by McCall et al. (\cite{mcca}), for all Balmer lines. A similar value
was found by D\'\i az et al. (\cite{di87}) in their analysis of NGC 604. In
most cases the correction is quite small and will be less than the quoted error
bars, except for H$\delta$ in the faintest regions and at H$\epsilon$, where it
may constitute a fair fraction of the measured line flux. We deal with this
problem by assigning errors to include no
correction and corrections up to 4 \AA~equivalent width in a fashion similar
to Puxley \& Brand (\cite{pu:br}). The large uncertainty for some of the
spectra further cautions the usage of H$\epsilon$.

\begin{table*}
\caption{\label{flux} Observed line fluxes (normalised at
${\cal F}({\rm H}\beta) = 1$)}
\begin{flushleft}
\begin{tabular}{lcccccccccccc}
\noalign{\smallskip}
\hline
\noalign{\smallskip}
\multicolumn{1}{c}{ID} & \multicolumn{12}{c}
{$R_{\rm o} = \left( \frac{{\cal F}(\lambda)}
 {{\cal F}({\rm H}\beta)} \right)_{\rm o}$} \\
\noalign{\smallskip\hrule\smallskip}
& N595 & B88,1 & B88,2 & N604,1 & N604,2 & N604,3 & N5461 & H949 & H961 &
H1026 & N5461 & H1117 \\
\noalign{\smallskip\hrule\smallskip}
H$\epsilon$ & 0.102 & 0.130 & 0.122 & 0.137 & 0.125 & 0.096 & 0.104 & 0.122 
& 0.098 & 0.091 & 0.101 & 0.188 \\
H$\delta$ & 0.232 & 0.235 & & 0.253 & 0.256 & 0.252 & 0.209 & 0.210 
& 0.212 & 0.178 & 0.191 & \\
H$\gamma$ & 0.409 & 0.437 & 0.491 & 0.441 & 0.456 & 0.451 & 0.392 & 0.402 
& 0.395 & 0.387 & 0.365 & 0.370 \\
H$\beta$ & 1.000 & 1.000 & 1.000 & 1.000 & 1.000 & 1.000 & 1.000 & 1.000 
& 1.000 & 1.000 & 1.000 & 1.000 \\
P14 & 0.011 & & & 0.010 & 0.009 & & 0.016 & & & \\
P13 & 0.019 & 0.009 & & 0.013 & 0.012 & & 0.020 & & & & 0.022 & \\
P12 & 0.023 & 0.014 & 0.013 & 0.014 & 0.015 & & 0.020 & & & & 0.034 & \\
P11 & 0.029 & 0.015 & 0.020 & 0.017 & 0.018 & & 0.025 & & & & 0.039 & \\
P10 & 0.048 & 0.022 & 0.044 & 0.021 & 0.021 & 0.029 & 0.035 & 
& 0.049 & 0.062 & 0.039 & 0.115 \\
P9  & 0.064 & 0.034 & 0.052 & 0.038 & 0.037 & 0.055 & 0.059 & 0.057
& 0.068& 0.110 & 0.071 & 0.126 \\
P$\delta$ & 0.159 & 0.085 & 0.164 & 0.090 & 0.101 & 0.064 & 0.158 & 0.145 
& 0.200 & 0.360 & 0.211 & \\
P$\gamma$ & & & & & 0.378 & & 0.297 & & & & & \\
\noalign{\smallskip}
\hline
\noalign{\smallskip}
${\cal F}\left({\rm H}\beta\right)^a$ & -12.55 & -13.07 & -13.48 & -12.43 &
-12.61 & -13.59 & -12.73 & -13.38 & -13.50 & -13.90 & -12.79 & -14.43 \\
\noalign{\smallskip}
\hline
\noalign{\smallskip}
\end{tabular}
\begin{list}{}{}
\item[$^{\rm a}$] Logarithm of the H$\beta$-flux in units of $\rm
erg~cm^{-2}~s^{-1}$\\
\end{list}
\end{flushleft}
\end{table*}

In Table \ref{flux} we list the line flux relative to H$\beta$ of all
identified hydrogen emission lines of the observed \hrs~ in NGC 598 and NGC
5457.

\section{Discussion}
\begin{figure*}
\parbox[b]{13cm}{\vspace{7.5cm}
\includegraphics{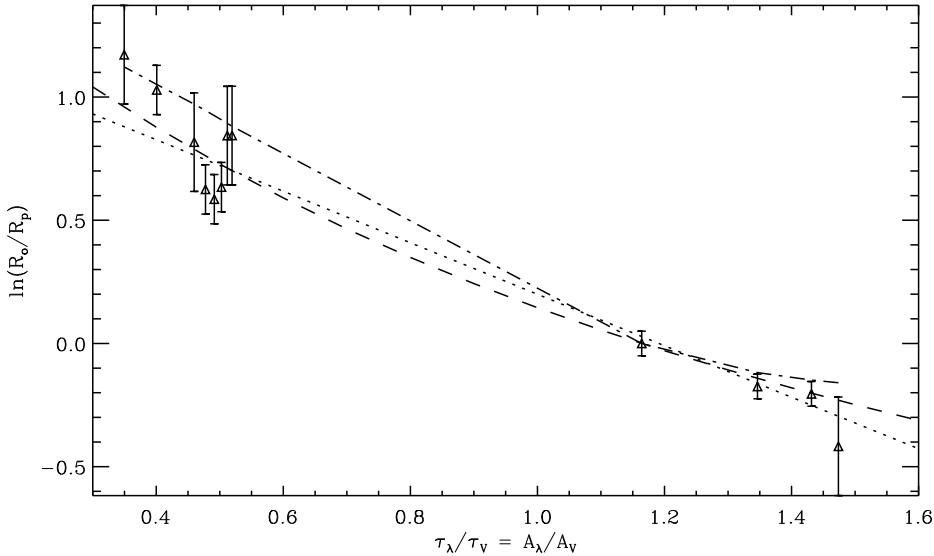}}
\hfill \parbox[b]{4.8cm}{\caption{\label{N5461}Observed and predicted line
flux relative to H$\beta$ in the \hr~ NGC5461 with three best fit models for
the dust distribution: Foreground screen (dotted line), homogeneous slab
(dashed line) and the WTC 'dusty nucleus' (dot-dashed)}}
\end{figure*}

\begin{figure*}
\parbox[b]{13cm}{\vspace{7.5cm}
\includegraphics{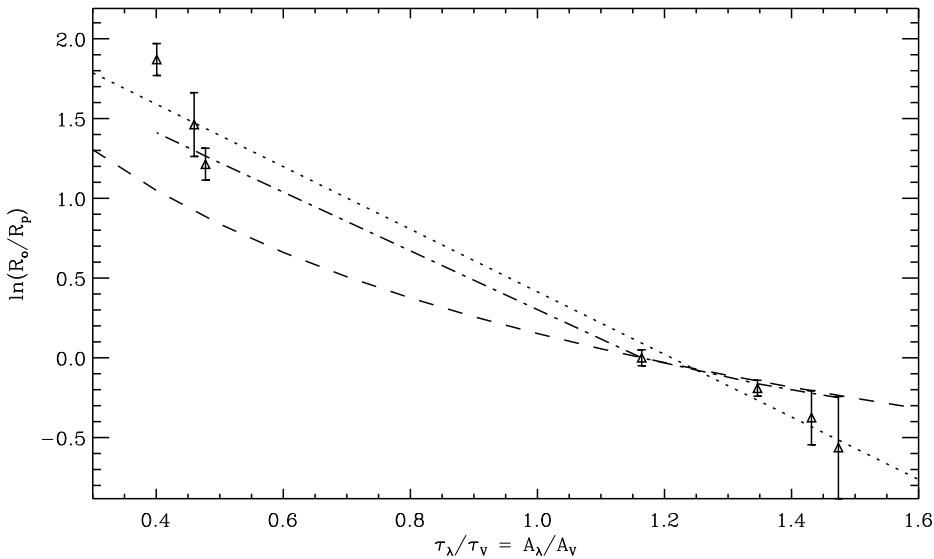}
}
\hfill \parbox[b]{4.8cm}{\caption{\label{H1026}Same as Fig. \ref{N5461} but for
the region H1026 where the quotients of observed and predicted line ratios
clearly cannot be fit by the homogeneous slab model
}}
\end{figure*}

\begin{figure*}
\parbox[b]{13cm}{\vspace{7.5cm}
\includegraphics{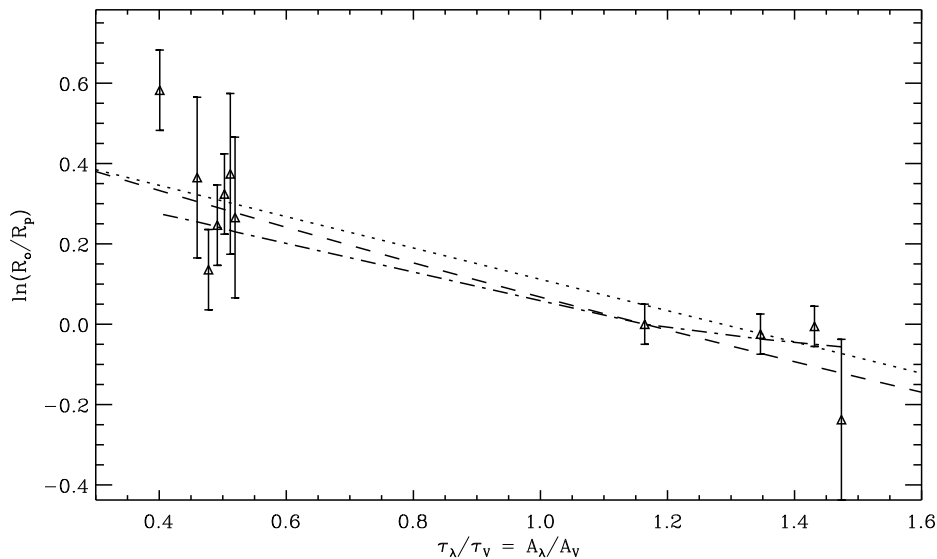}}
\hfill \parbox[b]{4.8cm}{\caption{As Fig. \ref{N5461} but for the central
emission region of the NGC 604 complex\label{N604}}}
\end{figure*}

The quotient of observed and predicted line fluxes in three of the observed
regions relative to H$\beta$,
\be
\frac{R_{\rm o}}{R_{\rm p}} = \left( \frac{{\cal F}(\lambda)}
{{\cal F}({\rm H}\beta)} \right)_{\rm o} \left/ \left( \frac{I(\lambda)}
{I({\rm H}\beta)} \right)_{\rm p} \right.
\ee
are plotted in Figs. \ref{N5461}--\ref{N604} with $\tau_{\lambda} / \tv = 
A_{\lambda} / A_{\rm V}$ on the abscissa as derived
from the fitting formula by Cardelli et al. (\cite{ccm}) with the standard
value $R_{\rm V} = 3.1$.
The choice of normalising to H$\beta$ is simply a matter of convenience and
when deriving $\tv$ from the corresponding multiplet lines the
relevant Balmer lines are used in the general expression Eq. (2).

\begin{table*}
\caption{Optical depth $\tv$ for the observed \hrs~ in NGC 5457 \label{N5457},
a '--' indicate that no acceptable fit could be obtained.
The Galactic extinction is 0. in this direction (de Vaucouleurs et al. 1991)
}
\begin{flushleft}
\begin{tabular}{llllllll}
\noalign{\smallskip}
\hline
\noalign{\smallskip}
 & \hfil N5461 & \hfil H949 & \hfil H961 & \hfil H1026 & \hfil N5461 & 
\hfil H1117 \\
\noalign{\smallskip}
\hline
\noalign{\smallskip}
Foreground screen & & & & & & \\
${\rm P}\gamma / {\rm H}\delta$ & $1.27 \pm 0.2$ & & & & & \\
${\rm P}\delta / {\rm H}\epsilon$ & $ 1.35 \pm 0.2 $& $1.14 \pm 0.3$ &
$1.65 \pm 0.2$ & $2.27 \pm 0.3$ & $1.65 \pm 0.2$ & \\
${\rm P}\delta / {\rm H}\beta$ & $1.35 \pm 0.2$ & $1.26 \pm 0.2$ & 
$1.68 \pm 0.2$ & $2.45 \pm 0.2$ & $ 1.73 \pm 0.2$ & \\
linear fit & $ 1.04 \pm 0.07 $ & $ 1.15 \pm 0.06 $ & $ 1.31 \pm 0.10 $
& $ 1.96 \pm 0.2 $ & $ 1.44 \pm 0.09 $ & $ 2.26 \pm 0.3 $ \\
Homogenous slab & 4.3 & 5.5 & -- & -- & -- & -- \\
Homogenous slab, eff & 1.5 & 1.7 & -- & -- & -- & -- \\
WTC & 3 & 3 & 3 & 4 & 3 & 5 \\
WTC, eff & 1.7 & 1.7 & 1.7 & 2.4 & 1.7 & 3.1 \\
\noalign{\smallskip}
\hline
\noalign{\smallskip}
Peimbert \& Spinrad (\cite{pe:sp}) & $0.40 \pm 0.1$ & & & & $0.40 \pm 0.10$ &\\
Smith (\cite{sm}) & 0.83 & & & & 0.83 & \\
Rayo et al. (\cite{ra82}) & 1.49 & \multicolumn{2}{c}
{( ~~~~~~~~ 1.39 ~~~~~~~~ )} & & 1.49 & \\
McCall et al. (\cite{mcca}) & $1.16 \pm 0.13$ & & & & $1.16 \pm 0.13$ & \\
Kennicutt \& Garnett (\cite{ke:ga}) & $0.71 \pm 0.10$ & $0.77 \pm 0.06$ & &
$0.22 \pm 0.04$ & $0.71 \pm 0.10$ & \\
Skillman \& Israel (\cite{sk:is})$^{\rm a}$& 1.23 & & & & 1.23 & \\
Rosa \& Benvenuti (\cite{ro:be})$^{\rm b}$& 0.87 & & & & 0.87 & \\
\noalign{\smallskip}
\hline
\noalign{\smallskip}
\end{tabular}
\begin{list}{}{}
\item[$^{\rm a}$]Derived from the Br$\gamma$/H$\beta$ ratio\\
\item[$^{\rm b}$]Based on HST observations and population models\\
\end{list}
\end{flushleft}
\end{table*}

The theoretical line ratios $R_{\rm p}$ for case B recombination
are adopted from Hummer \& Storey (\cite{hu:st}). The electron temperature
and density have been determined by Kwitter \& Aller (\cite{kw:al}), Rayo et
al. (\cite{ra82}), D\'\i az et al. (\cite{di87}) and V\'\i lchez et al.
(\cite{vi88}). For the two regions (H1026 and H1117) where $T_{\rm e}$ and
$n_{\rm e}$ can not be acquired from the literature we used typical values of
$10^4~{\rm K}$ and $10^2~{\rm cm^{-3}}$ and include realistic variations in the
error estimates in Tables 3 and 4. The weak variation of $R_{\rm p}$ with these
parameters will in any case only introduce uncertainties smaller than the
expected observational errors (Greve et al. \cite{gr94}).

According to Eq. (2) the overlying screen will manifest itself as a straight
line with a slope equal to $-\tv$ in this type of plot.
Figure \ref{N5461} illustrates this for NGC 5461.
The best linear fit
to the data points gives $\tv = 1.04$ (Table \ref{N5457}) with $\chi^2 = 13$,
but fails to match the data points at the longest wavelengths, i.e. smallest
$\tau_{\lambda} / \tv$. More complicated geometries introduce curvature in
the plots (see Eq. (3) and Figs. \ref{N5461}--\ref{H1026}). A homogeneous
slab with optical depth $\tv =
4.3$ fits the data equally well or perhaps even better ($\chi^2 = 11$),
especially at the long wavelength end. The 'dusty nucleus' model of $\tv =
3$ does not fit quite as well ($\chi^2 = 34$). Although it gives a good match
to P$\gamma$ and P$\delta$ it exaggerates somewhat the 'blueing' of the
spectrum at short wavelengths caused by scattering. Bearing the problems of the
H$\epsilon$ line flux in mind this might not be significant, though. 

Given the indicated uncertainties it would be wrong to rule out any of the
analysed dust configurations for this particular \hr, but it is clearly 
demonstrated that while the more
complex dust geometries have effective optical depths only moderately higher
than a foreground screen the true dust content can be as much 3--4 times 
higher. A definite discrimination of various models will need observations
of emission lines in the Paschen and Brackett series at still longer near-IR
wavelengths (Puxley \& Brand \cite{pu:br}).

From the observation of NGC 5461 in the other slit position the following night
we derive larger values of the optical depth, $\tv = 1.41$ for the foreground
screen and $\tv = 5$ for the 'dusty nucleus'. No acceptable fit was produced
by the homogeneous slab model. This deviation is larger than
the estimated uncertainties and we believe it is real and a consequence of the
non-coinciding slit positions, and only partly attributable to the effect of
differential refraction (Filippenko \cite{fil})
and the overall lower signal, since the principal Paschen lines all stand well
out of the noise level of the continuum and the slit is much larger than the
typical seeing encountered during the observations ($< 1\arcsec$) and was
close to the optimal orientation.
As discussed in section 4.1 variations of this order of magnitude are likely
to be present across giant \hrs~ at the spatial scale sampled here.

The optical depths for all the \hrs~ 
observed in NGC 5457 deduced from the three different dust distributions are
collected in Table \ref{N5457}.
H949 is the only other region where it was possible to obtain a best fitting
model from the homogeneous slab distribution with $\tv < 10$. (For larger
values of $\tv$ the slab model have an asymptotic behaviour which will not 
change the shape of the curve with increasing optical depth.) The data
from this \hr~ are also well fit with a 'dusty nucleus' model of $\tv = 3$.
For H1026 and H1117 we find remarkable high values of the optical depth 
that obviously cannot be fit by the homogeneous slab (see Fig. \ref{H1026}),
and is best described by a WTC model with $\tv$ equal 4 or 5, suggesting
that quite large amounts of dust are associated with the spiral arm along which
these two regions are positioned.

In Fig. \ref{N604} we plot the relative line fluxes of NGC 604 in
NGC 598 in the same fashion as for NGC 5461. Since P$\gamma$ is just barely
identified it
has not been included in the diagram. Evidently, this \hr~ experiences much
less dust extinction making it even harder to distinguish between any of the
three models. The plotted curves fit the data points equally well with both
the homogeneous slab model and the 'dusty nucleus' model having $\tv = 1$, but
apparently the scattering properties of the WTC model follow better the 
flattening of $\ln ( R_{\rm o}/ R_{\rm p})$ for the blue optical lines.
None of the models adequately fit the measurement of P$\delta$, which indicate
higher extinction.

Table \ref{N598} lists the values of the optical depth derived from the three
models for the observed \hrs~in NGC 598.
In general they show low values of $\tv$ with the highest found for NGC 595.
For this \hr~ the observations are best
described by a homogeneous slab or the 'dusty nucleus' model again with
indications of a 'blueing' of the spectrum at short wavelengths. For the 
other observations none of the considered models stand out.
\begin{table*}
\caption{Optical depth $\tv$ for the observed \hrs~ in NGC 598.
$\tau^{Gal}_{\rm V} = 0.13$ \label{N598} (de Vaucouleurs et al. 1991)}
\begin{flushleft}
\begin{tabular}{llllllll}
\noalign{\smallskip}
\hline
\noalign{\smallskip}
 & \hfil N595 & \hfil B88,1 & \hfil B88,2 & \hfil N604,1 & \hfil N604,2 &
\hfil N604,3 \\
\noalign{\smallskip}
\hline
\noalign{\smallskip}
Foreground screen & & & & & & \\
$ {\rm P}\gamma  / {\rm H}\delta $ & & & & & $ 1.31 \pm 0.4$ & \\
$ {\rm P}\delta / {\rm H}\epsilon$ & $1.38 \pm 0.2$ & $0.54 \pm 0.4$ &
$ 1.22 \pm 0.3$ & $0.57 \pm 0.2$ & $ 0.76 \pm 0.2$ & $ 0.59 \pm 0.3$  \\
$ {\rm P}\delta  / {\rm H}\beta $ & $1.35 \pm 0.2$ & $0.52 \pm 0.2$ & $1.38
\pm 0.2$ & $0.62 \pm 0.2$ & $0.76 \pm 0.2$ & $0.18 \pm 0.2$ \\
lin. fit & $1.08 \pm 0.08$ & $0.32 \pm 0.06$ & $0.87 \pm
0.15$ & $0.35 \pm 0.06$ & $0.39 \pm 0.09$ & $0.41 \pm 0.14$ \\
Homogenous slab & 6.5 & 0.8 & 3.8 & 0.9 & 1.0 & 1.0 \\
Homogenous slab, eff & 1.9 & 0.4 & 1.4 & 0.4 & 0.5 & 0.5 \\
WTC & 3 & 1 & 2 & 1 & 1 & 1 \\
WTC, eff & 1.7 & 0.5 & 1.1 & 0.5 & 0.5 & 0.5 \\
\noalign{\smallskip}
\hline
\noalign{\smallskip}
Peimbert \& Spinrad (\cite{pe:sp}) & & & & ( & $0.44 \pm 0.10$ & \hfill )
\\
Smith (\cite{sm}) & 1.00 & & & ( & 0. &  \hfill ) \\
French (\cite{fr}) & & & & ( & 0.80 & \hfill ) \\
Kwitter \& Aller (\cite{kw:al}) & 1.35 & \multicolumn{2}{c}
{( ~~~~~~~~ 0.50 ~~~~~~~~ )} & ( & 0.75 & \hfill ) \\
McCall et al. (\cite{mcca}) & 0.95 & & & & & & \\
D\'\i az et al. (\cite{di87}) & & & & 0.59 & 0.59 & 0.79 \\
Melnick et al. (\cite{mel87}) & 0.77 & & & ( & 0.59 & \hfill ) \\ 
V\'\i lchez et al. (\cite{vi88})&  & \multicolumn{2}{c}
{( ~~~~~~~~ 0.20 ~~~~~~~~ )} & ( & 0.69 & \hfill ) \\
\noalign{\smallskip}
\hline
\noalign{\smallskip}
\end{tabular}
\end{flushleft}
\end{table*}

Two pairs of corresponding multiplet lines, ${\rm P}\gamma / {\rm H}\delta$
and ${\rm P}\delta / {\rm H}\epsilon$, are included in the present observations
but the former only for two of the \hrs.
While the usage of any of the two pairs are not unproblematic and involve large
flux uncertainties they do on the other hand provide the advantage of common
upper levels and less reliance on the recombination models. Here they have
been included in Tables \ref{N5457} and \ref{N598} for reason of comparison
with derivations based on all observed emission lines.
It is evident that ratios of these Paschen and Balmer lines always yield a
higher value of $\tv$, although with larger errors than using the full set of
observed lines. This could
be due to an underestimation of the H$\epsilon$ flux in the deblending process
and the correction for underlying Balmer absorption or the low S/N
at the position of P$\gamma$ reflected in the quoted error bars. But Figs.
\ref{N5461} and \ref{N604} also show that P$\gamma$ and P$\delta$ always
lie above the straight line outlined by the remaining points, thereby
indicating some curvature away from the foreground screen model. This higher
optical depth determined from emission lines with long wavelength baselines
is further discussed in section \ref{other}.

Finally, we have included extinction calculations based on the single line pair 
made up of the Paschen and Balmer line with highest S/N, namely ${\rm P}\delta /
{\rm H}\beta$. The derived optical depths from this ratio agrees well with those
found on the basis of the corresponding line pair ${\rm P}\delta /
{\rm H}\epsilon$, expect in the case of N604,3 due to the very small H$\epsilon$
flux measured for this region.
For the faintest regions ${\rm P}\delta / {\rm H}\beta$ gives a smaller formal
uncertainty than the corresponding line pairs, but generally the uncertainties
are of the same size since ${\rm P}\delta / {\rm H}\beta$ has a somewhat
smaller separation. 

What changes to the derived physical properties of the \hrs~ does the choice of
extinction model make? This will depend on the wavelengths and strengths of
the relevant lines, but since the effective optical depths do not differ very
much among the models it is not expected to see large effects on derived
physical parameters. This can be illustrated by the oxygen abundance
calibrator $R_{\rm 23}$ = ([O{\sc\,ii}]~+~[O{\sc\,iii}])/H$\beta$ widely used
to determine the metallicity of spiral galaxies. For this ratio the main
difference will be on the [O{\sc\,ii}] lines due to the scattering included in 
the WTC model, while [O{\sc\,iii}] only will experience minute changes relative
to H$\beta$.

From the uncorrected line
fluxes of NGC 604 by Diaz et al. (\cite{di87}) we have calculated the oxygen
abundance using the calibration formula given in Zaritsky et al. (\cite{ZKH})
for the three cases of extinction models. Extinction correction assuming a
foreground screen and a homogeneous slab both give 12+log(O/H) = 8.80, while
the WTC model results in 8.82, i.e. virtually no change for the small dust
extinction
of NGC 604. The spectrum of NGC 5461 is affected by a larger dust extinction 
and from the data by Rayo et al. (\cite{ra82}) we find a difference in oxygen
abundance of 0.08 dex between the foreground screen and the 'dusty nucleus'
model. This is almost negligible compared to the estimated precision of the
oxygen abundance calibration, roughly 0.2 dex (Pagel et al. \cite{pag80}), 
although extinction correction by the WTC model will systematically infer higher
oxygen abundances.
This result makes the widespread use of a simple
foreground screen for dereddening justifiable at this magnitude of extinction
at least for lines not to far apart. For diagnostic lines with longer
separations larger changes are expected, for instance Bautista et al.
(\cite{bau}) find deviations of up to 12 \% in nebular abundances in a
revaluation of Orion Nebula data with an empirical effective nebular
extinction law.

\subsection{Spatial variation of the extinction}

\begin{figure}
\vspace{7.5 cm}
\includegraphics{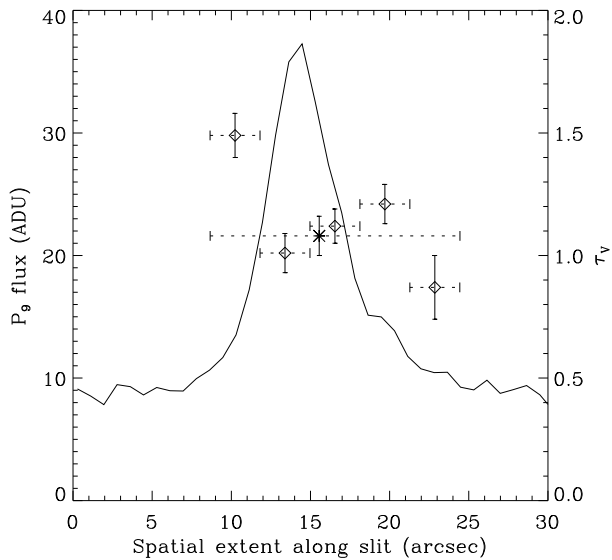}
\caption[]{\label{resolv} Spatially resolved optical depth across NGC 5461.
The solid line represents observed flux in the spatial direction along the
slit in ADU's on the left ordinate at the central wavelength of the P9 emission
line. The $\star$ denotes the optical
depth given on the right ordinate with associated vertical error bar as derived
from the total integrated spectrum. The size of the extraction
aperture is shown as the horizontal dotted line. The $\diamond$'s indicate
the optical depths found in the 5 subapertures extracted over the short
dotted lines}
\end{figure}

Two of the \hrs, NGC 5461 and NGC 604, has sufficiently high signal and large
extent that the spatial variation of the extinction across an \hr~ can be
studied in greater detail.
The E--W aperture (Jul 31) of NGC 5461 has been  divided into 5 subapertures of
length $3\farcs 2$ corresponding to 84 pc at the distance of NGC 5457 and the
extracted spectrum from each of them treated in the usual way.
In Fig. \ref{resolv} the optical depths derived in the large aperture and
each of the subapertures assuming the foreground geometry are depicted at their
positions along the slit on top of the observed flux in one of the Paschen
lines, which is less affected by the extinction.
As expected, the value of $\tv$ for the total aperture is close to those
found
in the central subapertures containing most of the flux with uncertainties
increasing outwards. A scatter in $\tv$ of $\sim 0.3$ across the region
is evident, which could indicate a patchiness of the obscuring dust
integrated along the line of sight of this order.
A similar behaviour is found by Greve et al. (\cite{gr89}) for the Galactic
Orion Nebula.

We also note that $\tv$ increases towards the edge of the region
(subaperture \#1 and 4). This can be interpreted as a larger amount of dust
residing just outside the brightest part of the \hr~ possibly due to destruction
and out-flow of dust in the inner starburst environment. It also fits well
with the picture upon which the WTC dusty nucleus simulations are based. Given
the uncertainty in
modelling the dust geometry discussed above one should, however, be cautious of
drawing any firm conclusion on the detailed distribution of the dust.

The separation of the three bright knots observed in the NGC 604 complex are
roughly 15\arcsec = 50 pc. From the overlying screen model we find here
somewhat smaller variations of $\tv$ up to 0.1, while Diaz et al.
(\cite{di87}) also found variations in $\tv$ of the order 0.3 among 5 areas
studied in detail. Thus there are signs of variation in the optical depths 
towards giant extragalactic \hrs~on scales smaller than 50 pc most likely
attributed to a patchy distribution of the obscuring dust.

\subsection{\label{other}Comparison with previous investigations}

The brightest \hrs~ in NGC 598 and NGC 5457 have been the objective of many
extensive studies over the years. In the lower part of Table \ref {N5457} and
\ref{N598} independent extinction determinations available in the literature
are compiled. The majority are based on the Balmer decrement
(H$\alpha$/H$\beta$), but for NGC 5461 two other sources are included. One is
derived from the ratio of P$\gamma$/H$\beta$, while Rosa \& Benvenuti
(\cite{ro:be}) use a
different approach fitting population synthesis models to the energy
distribution while taking the dust attenuation into account.

A substantial scatter between the different sources is seen, amounting to
more than one unit in optical depths for NGC 5461. Part of this can possibly be
explained by the different
aperture sizes and positions used if the dust distribution is not totally
homogeneous and entirely in the foreground, but it may also be related to the
problem of deriving the extinction from a pair of lines with only small
wavelength separation. The newer results seem to favour the generally high dust
extinctions that we derive, but also unexplainable wildly deviating values
exist.  

Our data display the same tendency as seen in Paper I and also commented by
Skillman \& Israel (\cite{sk:is}), that the derived extinctions increase with
increasing wavelength separation. 
A similar behaviour is found by Bautista et al. (\cite{bau}) in an analysis of
the Orion Nebula, where $A_{\rm V}$ estimated using the lines of a given
recombination line series (Balmer to Brackett) increases with the mean
wavelength of that line series. This finding is even based on a value of
$R_{\rm V} = 5.5$, which is appreciable altered from the value of 3.1 valid
for the interstellar medium. (With this value the differences would have been
more distinct).

This trend correlates with the findings by Viallefond \& Goss (\cite{vi:go} and
Skillman \& Israel (\cite{sk:is}) of yet higher values of extinction for 
NGC 598 and NGC 5457, respectively,
when comparing radio continuum emission and H$\beta$ line emission. Both 
attribute this 'excess' extinction to part of the dust being non-uniformly
distributed and/or mixed with the gas.
Skillman \& Israel (\cite{sk:is}) point out that most of the difference appears
to be explainable by dust associated with the \hrs, but there is no 
unambiguous method to distinguish between the two distributions.
It is interesting to see that the 
absorption at H$\beta$ for NGC 595 and NGC 5461 from these two studies when
converted to optical depth in V ($\tv =$ 1.4 and 1.8, respectively) match the
effective $\tv$ of the WTC 'dusty model' in our investigation.

Similarly Caplan \& Deharveng (\cite{ca:de}) report and average excess in
$A_{\rm V}$ of 0.3 mag for regions in the LMC and attribute this to clumped
interstellar dust and scattering dust in a slab near the periphery of the
emitting gas.
On the other hand Caplan et al. (\cite{cap96}) find in a recent study of the
SMC only a few regions with excess extinction and the reddening of the large
majority of \hrs~being consistent with uniform interstellar extinction. 
\section{Conclusions}

The observations of giant \hrs~ in the spiral galaxies NGC 598 of NGC 5457
have demonstrated how the dust extinction can be derived on the basis of
multiple emission lines from the Balmer and Paschen series in the optical and
near-IR regimes obtained simultaneously with ordinary CCD detectors. Although
some of these lines can only be observed with larger
flux errors the wide wavelength interval spanned by the lines and the
utilisation of the redundancy of the Paschen lines at
$\lambda\lambda$8500--9000 \AA~ leads to a small
overall uncertainty compared to the use of the two strongest Balmer lines. 

Inclusion of lines beyond 1 $\mu$m offers a chance to make simple test of
the geometrical distribution of emitting source and dust in the light of the
common but questionable assumption of a foreground screen. A slab of
homogeneous mixture can only match the data points for the \hrs~
with relative low dust extinction. Generally we can not discriminate between
the simple foreground extinction screen and the WTC 'dusty nucleus' model,
although signs of deviation from the pure foreground screen are seen at near-IR wavelengths. This is in agreement with Calzetti et
al. (\cite{CKS96}), who found that the reddening towards starburst regions
in 13 galaxies by large can be explained by foreground dust.
The choice of extinction model for reddening correction of diagnostic line
ratios only has small effects on the derived physical parameters when lines are
not widely separated.

It is, however, clearly
seen how the foreground screen always yields a lower limit to the dust content 
and that considerable larger amounts of dust can be present in the 'dusty
nucleus' configuration. A more definitive discrimination between various
geometries,
possibly including patchiness, and assessment of the actual dust content will
need investigations at even longer wavelengths in combination with optical
lines (Puxley \& Brand \cite{pu:br}; Genzel et al. \cite{gen95}; Calzetti et al.\cite{CKS96}) and more refined modelling (Witt \& Gordon \cite{wi96}).
While a lot can be learned from high-S/N observations of P$\gamma$ obtainable
with the high quantum efficiency of the new generation of CCD's, it is 
desirable to have data on emission lines further
out in the near-IR, e.g. P$\beta$, Br$\gamma$, Br$\alpha$, which can not be
observed with the same instruments as the blue optical lines. The kind of
studies presented in this paper can serve to bridge the two wavelength ranges.

One obvious result from the present study is the demonstration of a
spatial variation in the optical depth towards the emission regions of the
order 0.2--0.3 on small angular scales. More detailed investigations of one
of the \hrs~ suggest
that the extinction has it highest value at or close to the edge of the region.
This can explain at least some of the
disagreement in the quoted extinction of \hrs~ by different authors. It also
emphasizes the importance of sampling exactly the same slit position and
aperture
size if combining data from different observations, and the care that most be
taken when comparing physical parameters from miscellaneous sources in the
literature.

\end{document}